# Assessing the Performance of Deep Learning for Automated Gleason Grading in Prostate Cancer

Dominik MÜLLER[1,a,b], Philip MEYER[b], Lukas RENTSCHLER[b,c], Robin MANZ[b], Daniel HIEBER[c,d,e], Jonas BÄCKER[b], Samantha CRAMER[b], Christoph WENGENMAYR[b], Bruno MÄRKL[c], Ralf HUSS[c,f], Frank KRAMER[a], Iñaki SOTO-REY[b], Johannes RAFFLER[b,e]
[a]*Faculty of Applied Computer Science, University of Augsburg, Germany*
[b]*Institute for Digital Medicine, University Hospital Augsburg, Germany*
[c]*Institute for Pathology, University Hospital Augsburg, Germany*
[d]*Institute DigiHealth, Neu-Ulm University of Applied Sciences, Germany*
[e]*Bavarian Cancer Research Center (BZKF), Augsburg, Germany*
[f]*BioM Biotech Cluster Development GmbH, Germany*

**Abstract.** Prostate cancer is a dominant health concern calling for advanced diagnostic tools. Utilizing digital pathology and artificial intelligence, this study explores the potential of 11 deep neural network architectures for automated Gleason grading in prostate carcinoma focusing on comparing traditional and recent architectures. A standardized image classification pipeline, based on the AUCMEDI framework, facilitated robust evaluation using an in-house dataset consisting of 34,264 annotated tissue tiles. The results indicated varying sensitivity across architectures, with ConvNeXt demonstrating the strongest performance. Notably, newer architectures achieved superior performance, even though with challenges in differentiating closely related Gleason grades. The ConvNeXt model was capable of learning a balance between complexity and generalizability. Overall, this study lays the groundwork for enhanced Gleason grading systems, potentially improving diagnostic efficiency for prostate cancer.

## 1. Introduction

The ever-growing burden of prostate cancer, exceeding 68,000 diagnoses annually in Germany alone, necessitates the development of robust and efficient diagnostic tools [1]. The Gleason scoring system remains a cornerstone in guiding treatment decisions, which depends on the accurate identification of cancer in tissue sections [2]. Recent advancements in digital pathology and artificial intelligence, particularly deep neural networks, offer great potential for improved diagnostic robustness and efficiency [3]. These medical image analysis models have showcased their prowess in detecting complex patterns within digitized tissues, both established and novel ones, making them invaluable assets in aiding tumor diagnosis and biomarker prediction [3]. While established neural network architectures from a decade ago have paved the way, recent advances demonstrated cutting-edge architectures pushing the boundaries of computer vision [4]. However, existing Gleason grading research often overlooks this progress, heavily relying on older architectures.

To address this gap, this study delves into a comprehensive performance comparison, evaluating both traditional and cutting-edge deep neural network architectures for automatic Gleason grading via tile-based image classification of prostate carcinoma. This study aims to not only provide detailed insights into the capabilities of state-of-the-art automated Gleason grading pipelines but also pave the way for future advancements in this crucial area of cancer diagnosis.

## 2. Methods

In order to ensure comprehensive and reliable evaluation of various neural network architectures for Gleason grading, we established a standardized pipeline based on the AUCMEDI framework [5], specifically designed for building medical image classification pipelines. This approach facilitated seamless switching between architectures, enabling a robust comparison and detailed evaluation of their performance based on equal conditions.

### 2.1. Prostate Carcinoma Dataset

In a recent in-house retrospective study, we extracted 325 prostate cancer cases diagnosed between 2019 and 2021 within the University Hospital Augsburg. We digitalized and annotated 369 H&E-stained tissue slides obtained from these patients ourselves. This detailed annotation process involved identifying various tissue types and potential artifacts such as air pockets or tissue distortion. Subsequently, we divided these annotated whole-slide images into small patches to create a dataset for training and evaluating deep learning models. This division resulted in a dataset with 34,264 tiles which we sampled according to a 38-10-52 percentage split into training, validation, and testing subsets, respectively. For our analysis, this study incorporated five main tissue classes: regular tissue, representing normal prostate tissue (23.1%), and Gleason grades, which categorized cancer severity based on the Gleason scoring system. The data included Gleason 3 (mild cancer: 2.9%), Gleason 4 (moderately differentiated cancer: 8.4%), and Gleason 5 (poorly differentiated cancer: 7.5%). Additionally, we included two artifact classes: Artefact Empty, representing empty tissue including minor issues like dust or scratches (13.1%), and Artefact Sponge, indicating tissue distortion caused by sponges during biotechnical processing (45.0%). Examples of tiles are visualized in **Figure 1**.

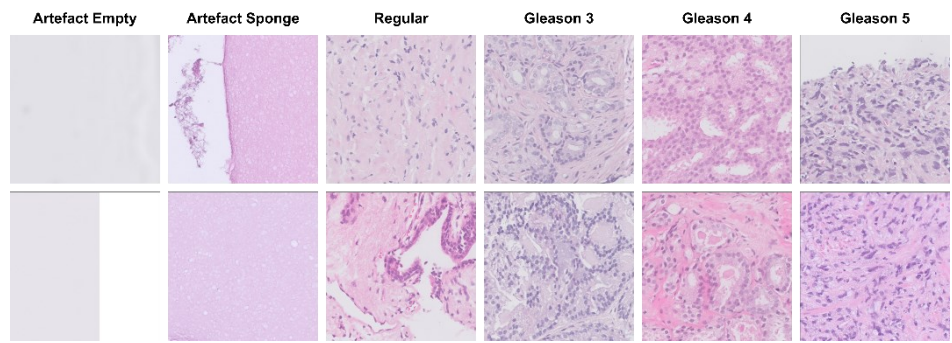

**Figure 1**. Visualization of the tile-based prostate carcinoma dataset with associated class examples.

*2.2. Image Preprocessing*

Before feeding images into the model, we applied various preprocessing techniques to enhance their quality and improve the model's pattern-recognition capabilities. During training, on-the-fly image augmentation diversified the dataset with flips, rotations, and color adjustments, avoiding artificial bias. Furthermore, preprocessing involved resizing tiles to match the model's input format (224x224 pixels), normalizing stain variations utilizing the method by Reinhard et al. [6], and standardizing pixel intensities across images utilizing Z-Score normalization. These steps prepared the data for optimal learning and neural network architecture comparison.

*2.3. Neural Network Models*

For the neural network model comparison, we deployed and evaluated a total of 11 deep learning architectures: VGG16, DenseNet121, ResNet101, MobileNetV2, ResNeXt101, Xception, InceptionV3, NASNet (variant Large), EfficientNet (variant B4), Vision Transformer (variant B16, short: ViT), and ConvNeXt (variant Base). Details and more information about the implementation can be found in the AUCMEDI documentation [5]. All models underwent a standardized training procedure to ensure comparability. Initially, a transfer learning approach fine-tuned the classification head for 10 epochs using pre-trained ImageNet weights and Adam optimization with an initial learning rate of $1e^{-4}$. Subsequently, fine-tuning encompassed the entire architecture for a maximum of 1,000 epochs, employing a dynamic learning rate starting at $1e^{-5}$ and progressively decreasing to $1e^{-7}$ (reduction factor of 0.1 after 5 epochs of validation loss plateau). Early stopping halted training after 10 epochs with no validation loss improvement. Training used a batch size of 28 samples, the traditional epoch definition, and a weighted Focal loss for optimization.

## 3. Results

We were able to successfully build a standardized pipeline as foundation for comparing the performance of various deep learning architectures in Gleason grading for prostate cancer classification. The performance of all architectures is illustrated in **Figure 2**.

We observed that sensitivity for the different architectures significantly varied, with ConvNeXt achieving the highest sensitivity at 83%, followed by VGG16 at 80%. In contrast, EfficientNet demonstrated the lowest sensitivity at 68%. Notably, the ViT architecture ranked third-last in sensitivity at 75%, slightly behind the minimalistic architecture MobileNetV2 with 76% sensitivity. All architectures demonstrated a consistently high specificity, averaging around 97%. The overall accuracy of the models ranged from 94.5% to 97.5%, with a mean accuracy of 96.5%. Notably, the accuracy results were consistent with the sensitivity order across all architectures.

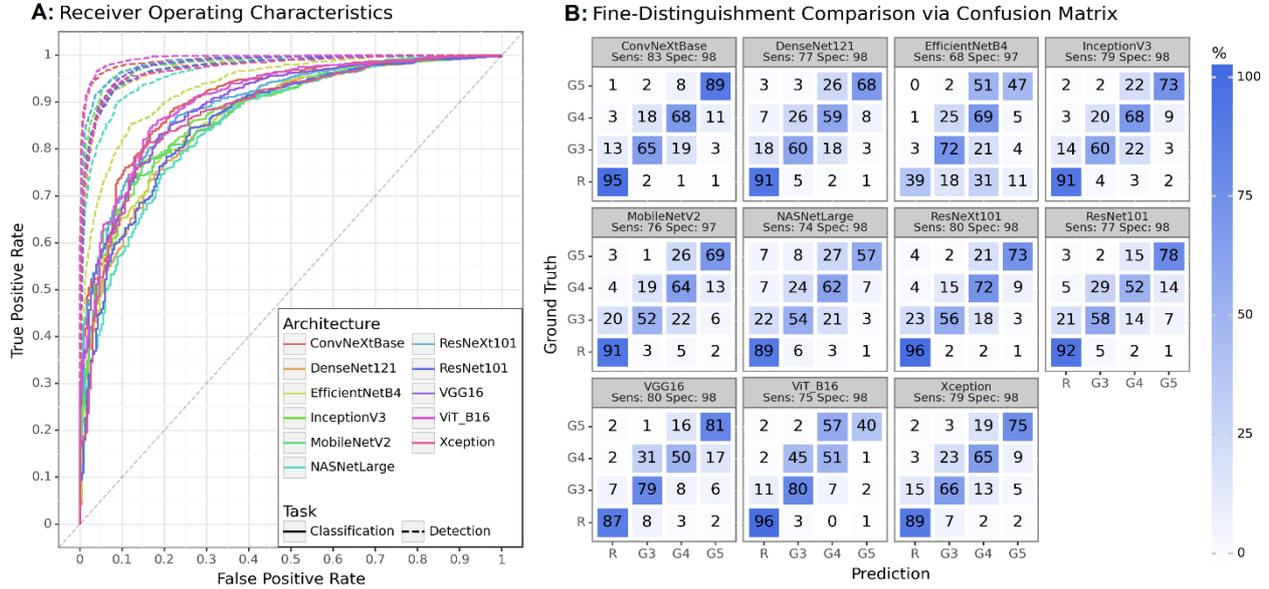

**Figure 2.** Comparison of deep neural network architectures for Gleason grading. **A:** Receiver Operating Characteristics (ROC) curves categorized into the detection and classification task. **B:** Confusion matrix between predicted and annotated classes illustrating the overall performance and showing the Sensitivity and Specificity.

Furthermore, we evaluated the performance in two popular categories commonly utilized in the literature to assess Gleason grading performance: detection and classification. In the detection task, the objective is to distinguish between malignant (Gleason grades G3-G5) and benign tissue (regular tissue). EfficientNet demonstrated the highest sensitivity at 99%, indicating its ability to accurately detect malignant tissue. However, it also exhibited the lowest specificity at 28%, suggesting a higher rate of false positives. Conversely, ConvNeXt achieved a sensitivity of 96% with a higher specificity of 94%, indicating its effectiveness in correctly identifying malignant tissue while minimizing false positives. In the classification task, the aim is to assess fine distinction within cancer severity by distinguishing between Gleason grade 3 (G3) and higher grades (G4-G5). ResNeXt101 and ConvNeXt showed the highest sensitivity at 89% and 88%, respectively, indicating their capability to accurately classify tissue into different Gleason grades. However, ResNeXt101 had a specificity of 67%, indicating a higher rate of false positives, whereas ConvNeXt exhibited a higher specificity of 75%. On the other hand, ViT demonstrated the lowest sensitivity of 72% but compensated with the highest specificity of 87%, suggesting its ability to minimize false positives at the expense of missing positive cases.

## 4. Discussion

Gleason grading of prostate cancer presents a multifaceted problem due to the substantial intricacy of histopathological pattern interpretation [2]–[4]. Heterogeneous tumor morphologies characterized by coexisting architectural and cellular features further complicate this task [2], [4]. Discerning these subtle nuances requires expertise in both normal and pathological tissue structures. Our proposed pipeline, utilizing various deep learning architectures, demonstrates remarkable performance in automated Gleason grading. Not only does it replicate previous studies' results, but it surpasses them by employing cutting-edge architectures for enhanced reliability [3], [4].

Comparing newer architectures like ViT and ConvNeXt to established models like DenseNet and ResNet yielded noteworthy observations. As anticipated, newer architectures exhibited superior performance, likely due to their increased complexity and capacity to capture intricate patterns more effectively, especially for differentiating closely related Gleason grades. However, this increased capacity can lead to overfitting on small datasets, hindering generalization to unseen data. We suspect this occurred with the EfficientNet and ViT model, as complexity-scaled convolutional neural networks as well as transformers often require vast and diverse training data. Conversely, the ConvNeXt model struck a balance between complexity and generalizability.

Deviations from ground truth primarily involve adjacent classes, reflecting the ordinal structure of the Gleason system. The models struggled most with differentiating regular, Gleason 3, and Gleason 4 tissues which is why the distinguishment between these subtly transitioning classes based on the ordinal nature of Gleason grades demonstrated the dominant aspect for our architecture comparison. This challenge, inherent to the ordinal nature of Gleason grades, also manifests in our model's predictions.

## 5. Conclusions

This study compared established and cutting-edge deep learning architectures for automated Gleason grading. Newer architectures, particularly ConvNeXt, showed superior performance, demonstrating the benefits of incorporating recent advancements in computer vision. Nevertheless, challenges remain, especially with differentiating similar grades, highlighting the need for further research towards robustness and generalizability. This work paves the way for future development and potential clinical integration of automated Gleason grading systems, potentially improving the robustness of the Gleason scoring and diagnostic efficiency in prostate cancer.